\documentclass[fleqn,usenatbib]{mnras}

\usepackage{newtxtext,newtxmath}

\usepackage[T1]{fontenc}
\usepackage{ae,aecompl}

\usepackage{graphicx}	
\usepackage{amsmath}	
\usepackage{amssymb}	
\usepackage{lastpage}
\newcommand{\angstrom}{\mbox{\normalfont\AA}}

\title[Internal Kinematics of M10 and M71]{Internal Kinematics of M10 and M71}

\author[N. Barth et al.]{Nicholas A. Barth,$^{1,3,4}$\thanks{nbarth@ufl.edu}
Jeffrey M. Gerber,$^{2,3,4}$
Owen M. Boberg,$^{3}$ \newauthor
Eileen D. Friel,$^{3,4}$
and Enrico Vesperini$^{3}$
\\
$^{1}$Astronomy Department, University of Florida, Bryant Space Science Center 401, 1772 Stadium Rd, Gainesville, FL 32611, USA\\
$^{2}$Max Planck Institut f{\"u}r Astronomie, K{\"o}nigstuhl 17, 69117, Heidelberg, Germany\\
$^{3}$Astronomy Department, Indiana University Bloomington, Swain West 319, 727 East 3rd Street, Bloomington, IN 47405-7105, USA\\
$^{4}$Visiting astronomer, Kitt Peak National Observatory, National Optical Astronomy Observatory, which is operated by the \\ Association of Universities for Research in Astronomy (AURA) under a cooperative agreement with the National Science Foundation.
}

\date{Accepted 2020 April 8. Received 2020 April 8; in original form 2019 June 13}

\pubyear{2020}

\begin{document}
\label{firstpage}
\pagerange{\pageref{firstpage}--\pageref{lastpage}}
\maketitle

\begin{abstract}
We present a study of the internal kinematics of two globular clusters, M10 (NGC 6254) and M71 (NGC 6838), using individual radial velocity (RV) measurements obtained from observations using the Hydra multi-object spectrograph on the WIYN 3.5 m telescope. We measured 120 RVs for stars in M10, of which 107 were determined to be cluster members. In M71, we measured 82 RVs and determined 78 of those measurements belonged to cluster members. Using the cluster members, we determine a mean RV of $75.9 \pm 4.0$ (s.d.) km s$^{-1}$ and $-22.9 \pm 2.2$ (s.d.) km s$^{-1}$ for M10 and M71, respectively. We combined the Hydra RV measurements with literature samples and performed a line-of-sight rotational analysis on both clusters. Our analysis has not revealed a statistically significant rotation in either of these clusters with the exception of the inner region (10\arcsec - 117\arcsec) of M10 for which we find hints of a marginally significant rotation with amplitude V$_{rot} = 1.14 \pm 0.18$ km s$^{-1}$. For M10, we calculate a central velocity dispersion of $\sigma_{0} = 5.44 \pm 0.61$ km s$^{-1}$, which gives a ratio of the amplitude of rotation to the central velocity dispersion $V_{rot}/\sigma_{0} = 0.21 \pm 0.04$. We also explored the rotation of the multiple stellar populations identified in M10 and M71 and found rotation (or lack thereof) in each population consistent with each other and the cluster global rotation signatures.
\end{abstract}




\section{Introduction}

A growing number of observational studies based on large radial velocity surveys and proper motion measurements from the Hubble Space Telescope and the Gaia mission are shedding light on the internal kinematics of globular clusters and revealing a complex dynamical picture characterized by the presence of internal rotation and anisotropy in the velocity distribution (see e.g. \citealt{lane2011}, \citealt{bellazzini}, \citealt{fabricius}, \citealt{watkins}, \citealt{kimmig}, \citealt{gaia}, \citealt{kamann}, \citealt{ferraro}, \citealt{bianchini2018}, \citealt{jindal}, \citealt{sollima}).

On the theoretical side, a number of studies of the dynamical evolution of globular clusters have clearly shown that a complete characterization of the phase space of these stellar systems is essential to understand the role played by internal dynamical processes and the interaction with the external tidal field of the host galaxy in shaping their current observed properties. In particular, the results of numerical and theoretical investigations have shown that the extent of radial anisotropy in the velocity distribution is linked to the cluster's initial structure and its mass loss history (see e.g. \citealt{tiongco2016}) and that as a cluster evolves, angular momentum redistribution and loss due to escaping stars imply that the present-day internal rotation is a lower limit on the amount of initial rotation at the time of the cluster's formation (see e.g. \citealt{einsel}, \citealt{ernst}, \citealt{hong}, \citealt{tiongco}, \citealt{tiongco2019}, and references therein).

The discovery of multiple stellar populations distinguished by differing abundance patterns of light elements such as C, N, O, Na, O, Mg, and Al (see e.g. \citealt{gratton}, \citealt{gratton2019} and references therein) has raised new questions in the study of globular clusters and added another complex layer to their characterization. The kinematic properties of multiple populations can contain key information on their formation and dynamical history and a few initial observational studies have revealed interesting differences in the kinematics of the multiple stellar populations. \citet{richer}, \citet{cordero2015}, \citet{bellini}, \citet{bellini2018}, \citet{milone}, and \citet{cordoni} used proper motion measurements based on, respectively, HST and Gaia data to show that second-generation stars are characterized by a more radially anisotropic velocity distribution than first-generation stars, a result consistent with the expected kinematic evolution if second-generation stars formed more centrally concentrated than the first-generation population. \citet{cordero2017} studied the rotational properties of the multiple populations of M13 and found a stronger rotation for the extreme second generation population (see e.g. \citealt{bekki}, \citealt{bekki2011} for simulations predicting a stronger rotation for second-generation stars forming in a rotating first-generation cluster). An interesting difference between the velocity dispersion of first- and second-generation stars possibly due to differences in the properties of binary stars of the two populations has been reported recently by \citet{dalessandro}.

Differences in the phase space properties of multiple populations can be gradually erased during the cluster's long-term dynamical evolution and indeed a few studies (see e.g. \citealt{bellazzini}, \citealt{libralato}) have found clusters in which multiple stellar populations are characterized by similar spatial and kinematic properties. It is important to point out, however, that stronger differences may be present in the clusters' outermost regions which are not included in most observational studies. Wide-field studies of globular clusters' spatial and kinematic properties spanning a broad range of radial distances will be needed to build a complete picture of the dynamical properties of multiple populations.

In an effort to further explore the rotation properties of globular clusters and their multiple populations we present here results of a study of M10 (NGC 6254) and M71 (NGC 6838). The rotation of M10 was studied by \citet{ferraro} as part of the MIKiS survey, and both clusters were studied in the Gaia survey for internal rotation using proper motions \citep{sollima}. The results from \citet{sollima} do not reveal any rotation in M10 and find a $2\sigma$ rotation detection in M71 (see \citealt{sollima} for a summary of all the studies of rotation in M10 and M71 and other clusters). Both M10 and M71 have been studied for the properties of their multiple populations (\citealt{carrettaB}, \citealt{gerber}, \& \citealt{gerber2020}).

The two GCs have other properties that make them interesting objects to study. M10 is a nearby halo GC with a disk-like motion around the galaxy \citep{chen}. M71, on the other hand, is a well-studied, metal-rich GC with a low Galactic latitude \citep{cudworth}. Gaia data analyzed by \citet{baumgardt2019} indicate that orbits for both clusters are confined within the solar circle, with M10 reaching a radial distance of only 4.6 kpc, while M71 orbits between $\sim$5 to 7 kpc from the Galactic center. A study of their kinematics allows for insight into how the Galaxy's tidal field affects their structural evolution (see e.g. \citealt{cadelano} for a study suggesting that M71 is likely to have lost a large fraction of its initial mass).

This paper is organized as follows. In Section \ref{Observations} we discuss our observational data and radial velocity (RV) measurements, including the literature sources from which we obtained additional RV measurements for M10 and M71. The procedure followed to determine the rotational properties and the results obtained for each cluster are described in Section \ref{Analysis}. Finally, in Section \ref{Conclusions} we offer final thoughts and discuss future work to continue research in GC kinematics.

\section{Observations and Data Reduction} \label{Observations}

\subsection{Observations and Target Selection}

Using the Hydra multi-object spectrograph at the Wisconsin-Indiana-Yale-NOAO\footnote{The WIYN Observatory is a joint facility of the University of Wisconsin-Madison, Indiana University, the National Optical Astronomy Observatory, University of Missouri, and Purdue University.} (WIYN) 3.5 m telescope, we obtained observations of stars in M10 and M71 on two separate observing runs: 24 Jun. 2016 and 11-13 Jun. 2017. We used echelle grating ``316@63.4" for a dispersion of 0.16 \r{A} pix$^{-1}$ (7.7 km s$^{-1}$ pix$^{-1}$), over the wavelength range from 6030 \r{A} to 6350 \r{A}.

One of the main goals of this work was to explore any possible differences in the kinematics between the two populations that are known to exist in each cluster (\citealt{cordero2015}, \citealt{carrettaA}, \citealt{carrettaB}, \citealt{gerber}, \citealt{bowman}, \citealt{gerber2020}). While the radial velocities measured by \citet{carrettaA} and \citet{carrettaB} in M10 come from high resolution spectroscopy and are precise enough to use to determine population kinematics such as rotation, the studies by \citet{gerber} and \citet{gerber2020} used low resolution spectroscopy to classify stars based on molecular CN band features. These low resolution radial velocities do not have the precision necessary to measure the relatively small rotation signatures observed in GCs. Therefore, our main sample focused on observing stars in M10 and M71 that had classification information from \citet{gerber} and \citet{gerber2020}, but required high resolution follow up to determine more precise radial velocities.

In addition, we measured 21 stars in common with \citet{carrettaA} and \citet{carrettaB} from M10 and 11, 12, and 5 stars in common with various studies from M71 (\citealt{carrettaA}, \citealt{cordero2015}, \citealt{cohen}, respectively) to determine any possible systematic offsets in radial velocities measured by either study. If there were any unused fibers in a configuration, we placed them on stars that were previously unmeasured in the literature. Even though these stars did not have population information, they could be used to increase our sample size for our determination of the overall rotation of each cluster.

The sample of stars in each cluster covers the full extent of the red giant branch (similar to \citealt{gerber} and \citealt{gerber2020}), with a faint limit of a V magnitude of 18. To prevent bright stars from over-saturating, we divided our configurations into bright and faint configurations. For M10, we defined a V magnitude of $15 < V < 17.5$ for the faint configuration and a V magnitude of $11.5 < V < 15$ for bright. Similarly, in M71, the faint configuration was between $15 < V < 18$ and the bright configuration was $12 < V < 16$. Faint configurations were exposed for a total integrated exposure time of 20-30 minutes. Bright configurations were exposed for a total integrated exposure time of 10 minutes. These total exposure times were designed to reach a minimum signal-to-noise ratio of approximately 15 for all stars in our sample. In total, we observed 141 stars in the field of M10 and 108 stars in the M71 field.

\subsection{Data Reduction and Radial Velocities} \label{datareduction}

The data were reduced using the Image Reduction and Analysis Facility (IRAF)\footnote{IRAF is distributed by the National Optical Astronomy Observatories, which is operated by the Association of Universities for Research in Astronomy, Inc., under cooperative agreement with the National Science Foundation.} and the \textit{dohydra} task. Flat fielding, wavelength calibration with a ThAr comparison lamp spectrum, and extraction of one-dimensional spectra were all performed with \textit{dohydra}. Multiple exposures of a star's spectrum were combined into a single spectrum that was used for RV measurements. RV standards were also observed on both the 2016 and 2017 observing runs (HD182488, HD194071, and HD212943). The one-dimensional spectra and RV standards were then cross-correlated with the IRAF package \textit{fxcor}.

In M10, we were unable to find a reliable RV measurement with \textit{fxcor} for 18 stars, due to low signal-to-noise or other problems with the spectra. To identify radial velocity members, we then divided the remaining stars into radial bins that contain a similar number of stars, formed the mean velocity in that bin, and removed stars that fell outside of 3$\sigma$ from the mean. $\sigma$ values in individual bins were similar to the $\sigma$ about the overall mean RV of the cluster. This process excluded an additional 13 stars from the rotational analysis. Finally, in the case of multiple RV measurements for a single star, we averaged them for a final RV value. M10 had 3 duplicate measurements that were within the $\sigma$ cut. In the end, we obtained RV measurements for 107 member stars in M10 with a mean RV of $75.9 \pm 4.0$ (s.d.) km s$^{-1}$.

In M71, we were unable to determine reliable RV measurements for a total of 22 stars. We followed the same procedure as in M10, where we excluded 4 stars from the rotational analysis because they fell outside of a 3$\sigma$ range in radial bins. The 3 duplicate measurements were combined yielding a total of 78 stars assumed to be members of M71, with a mean RV of $-22.9 \pm 2.2$ (s.d.) km s$^{-1}$.

The benefit of multiple exposures and observation runs was the presence of multiple measurements of stars to gauge the uncertainty of our RV measurements. Each star was cross-correlated against three RV standards, leading to three measurements of RV and RV uncertainties. The typical standard deviation among the RV measurements was 0.5 km s$^{-1}$, and we adopt that standard deviation as the individual error associated with each measurement. The typical \textit{fxcor} errors were larger by almost a factor of two, which is consistent with \citet{cote1994} and \citet{reijns}. Across the two observation runs, the rms of the $\Delta$RV between multiple measurements was 1.2 km s$^{-1}$. While larger than the 0.5 km s$^{-1}$ uncertainty, this higher rms is likely a result of the magnitudes of the stars with multiple measurements, which place them near the faint limit of our bright configurations, causing them to have a slightly lower S/N and higher measurement uncertainty than typical. The higher rms is also heavily influenced by one of these stars, which had a 2 km s$^{-1}$ difference between the 2016 and 2017 run. The other RV measurements have a $<$1 km s$^{-1}$ difference.

Figures \ref{m10rRV} and \ref{m71rRV} show the RV distribution of stars in the two clusters as a function of radial distance from the cluster center. A list of objects observed with coordinates, photometry, measured heliocentric RVs, RV errors, and membership information can be found in Table \ref{m10hd} and Table \ref{m71hd} for M10 and M71, respectively.

\begin{table*}
	\centering
	\caption{\textbf{M10 Hydra Data} 1. \citet{photometry} 2. Populations are numbered as (1): First Generation stars, (2): Second Generation Stars, (0): No information about the star's population (\citealt{gerber}, \citealt{carrettaA}, \citealt{carrettaB}). The complete table is available online.}
	\label{m10hd}
	\begin{tabular}{ccccccccc} 
		\hline
		\hline
		\textbf{ID$^{1}$} & \textbf{RA$^{1}$} & \textbf{DEC$^{1}$} & \textbf{V$^{1}$} & \textbf{B$-$V$^{1}$} & \textbf{RV} & \textbf{RV Error} & \textbf{Pop.$^{2}$} & \textbf{Membership}\\
        & (J2000) & (J2000) & (mag) & (mag) & (km s$^{-1}$) & (km s$^{-1}$)\\
		\hline
        9 & 254.29483 & -4.07889 & 11.85 & 1.52 & 81.5 & 0.36 & 2 & y\\
        \hline
        11 & 254.25158 & -4.10039 & 11.88 & 1.63 & 75.6 & 0.17 & 1 & y\\
        \hline
        14 & 254.38488 & -4.05022 & 12.00 & 1.61 & 78.3 & 0.12 & 2 & y\\
        \hline
        15 & 254.28967 & -4.12286 & 12.01 & 1.45 & 85.0 & 0.32 & 2 & y\\
        \hline
        26 & 254.27817 & -4.05294 & 12.42 & 1.17 & 70.1 & 0.29 & 2 & y\\
        \hline
        28 & 254.32108 & -4.22408 & 12.46 & 1.48 & 74.7 & 0.23 & 2 & y\\
        \hline
        33 & 254.39517 & -4.11894 & 12.60 & 1.42 & 76.7 & 0.18 & 1 & y\\
        \hline
        36 & 254.31013 & -4.18647 & 12.65 & 1.42 & 77.0 & 0.21 & 1 & y\\
        \hline
        43 & 254.26158 & -4.17247 & 12.78 & 1.36 & 73.2 & 0.15 & 1 & y\\
        \hline
        45 & 254.26938 & -4.10947 & 12.79 & 1.39 & 78.7 & 0.16 & 2 & y\\
        \hline
        47 & 254.15200 & -4.23894 & 12.96 & 1.23 & 115.1 & 0.60 & 0 & n\\
        \hline
        49 & 254.30721 & -4.14314 & 12.84 & 1.10 & 77.2 & 0.32 & 1 & y\\
        \hline
        63 & 254.29658 & -4.05533 & 13.05 & 1.01 & 74.1 & 0.29 & 2 & y\\
        \hline
        65 & 254.30013 & -4.10028 & 13.07 & 1.16 & 75.9 & 0.17 & 1 & y\\
        \hline
        73 & 254.35779 & -4.10489 & 13.19 & 1.31 & 73.2 & 0.19 & 1 & y\\
        \hline
        83 & 254.22775 & -4.14875 & 13.35 & 1.22 & 82.9 & 0.32 & 1 & y\\
        \hline
        84 & 254.23963 & -4.11397 & 13.35 & 1.23 & 75.7 & 0.19 & 2 & y\\
        \hline
        88 & 254.22646 & -4.09336 & 13.41 & 1.23 & 75.8 & 0.13 & 2 & y\\
        \hline
	\end{tabular}
\end{table*}

\begin{table*}
	\centering
	\caption{\textbf{M71 Hydra Data} 1. \citet{cudworth}. 2. Populations are numbered as (1): First Generation stars, (2): Second Generation Stars, (0): No information about the star's population \citep{gerber2020}. The complete table is available online.}
	\label{m71hd}
	\begin{tabular}{ccccccccc} 
		\hline
		\hline
		{\textbf{ID$^{1}$}} & {\textbf{RA$^{1}$}} & {\textbf{DEC$^{1}$}} & {\textbf{V$^{1}$}} & {\textbf{B$-$V$^{1}$}} & {\textbf{RV}} & {\textbf{RV Error}} & {\textbf{Pop.$^{2}$}} & {\textbf{Membership}}\\
        & {(J2000)} & {(J2000)} & {(mag)} & {(mag)} & {(km s$^{-1}$)} & {(km s$^{-1}$)}\\
		\hline
		1-102 & 298.43729 & 18.77303 & 14.33 & 1.09 & -21.4 & 0.25 & 2 & y\\
        \hline
        1-106 & 298.43329 & 18.77564 & 14.36 & 0.89 & -22.3 & 0.60 & 1 & y\\
        \hline
        1-111 & 298.40171 & 18.77565 & 14.89 & 1.32 & -22.3 & 0.50 & 2 & y\\
        \hline
        1-113 & 298.44063 & 18.77565 & 12.43 & 1.80 & -22.1 & 0.50 & 1 & y\\
        \hline
        1-18 & 298.46617 & 18.76811 & 14.43 & 1.07 & -24.9 & 0.41 & 1 & y\\
        \hline
        1-19 & 298.46308 & 18.76964 & 14.40 & 1.06 & -24.1 & 0.31 & 2 & y\\
        \hline
        1-2 & 298.46433 & 18.75703 & 14.79 & 1.23 & -24.8 & 0.50 & 1 & y\\
        \hline
        1-23 & 298.47258 & 18.77464 & 15.03 & 1.22 & -20.8 & 1.03 & 1 & y\\
        \hline
        1-34 & 298.45779 & 18.78342 & 14.45 & 1.05 & -25.7 & 0.60 & 2 & y\\
        \hline
        1-38 & 298.45396 & 18.78717 & 15.06 & 1.24 & -51.9 & 0.47 & 0 & n\\
        \hline
        1-42 & 298.44733 & 18.78789 & 14.22 & 1.02 & -20.4 & 0.60 & 2 & y\\
        \hline
        1-44 & 298.44646 & 18.80044 & 13.44 & 1.34 & -20.9 & 0.37 & 1 & y\\
        \hline
        1-45 & 298.45113 & 18.80061 & 12.36 & 1.76 & -19.9 & 0.50 & 2 & y\\
        \hline
        1-46 & 298.46463 & 18.80172 & 12.29 & 1.75 & -24.6 & 0.50 & 1 & y\\
        \hline
        1-48 & 298.48175 & 18.80239 & 14.39 & 1.07 & -24.1 & 0.28 & 2 & y\\
        \hline
        1-53 & 298.46096 & 18.81881 & 12.97 & 1.61 & -24.8 & 0.50 & 2 & y\\
        \hline
        1-54 & 298.45158 & 18.81258 & 14.49 & 1.03 & -23.3 & 0.30 & 1 & y\\
        \hline
        1-56 & 298.45117 & 18.80712 & 13.14 & 1.38 & -21.3 & 0.42 & 2 & y\\
        \hline
	\end{tabular}
\end{table*}

\subsection{Literature Studies}
\label{literature}

As shown in many studies (see e.g. \citealt{ferraro}) an accurate characterization of the rotational properties of globular clusters requires data for hundreds of stars; in order to maximize the number of stars used in our analysis, we combined our data with published radial velocity studies for M10 and M71. This increased our ability to detect rotation within the two clusters, and also minimized the chance of finding false signals of rotation from small sample sizes. Since a major goal was to study the similarities and differences between the rotational properties of the two populations that have been identified in each cluster (\citealt{cordero2015}, \citealt{carrettaA}, \citealt{carrettaB}, \citealt{gerber}, \citealt{bowman}, \citealt{gerber2020}), we limited our selection of literature samples only to those that include information on the multiple populations within M10 and M71.

\subsubsection{M10 Literature Values}

\begin{figure}
\centering
\includegraphics[trim = 0.4cm 0.4cm 0.4cm 0.4cm, scale=0.55, clip=False]{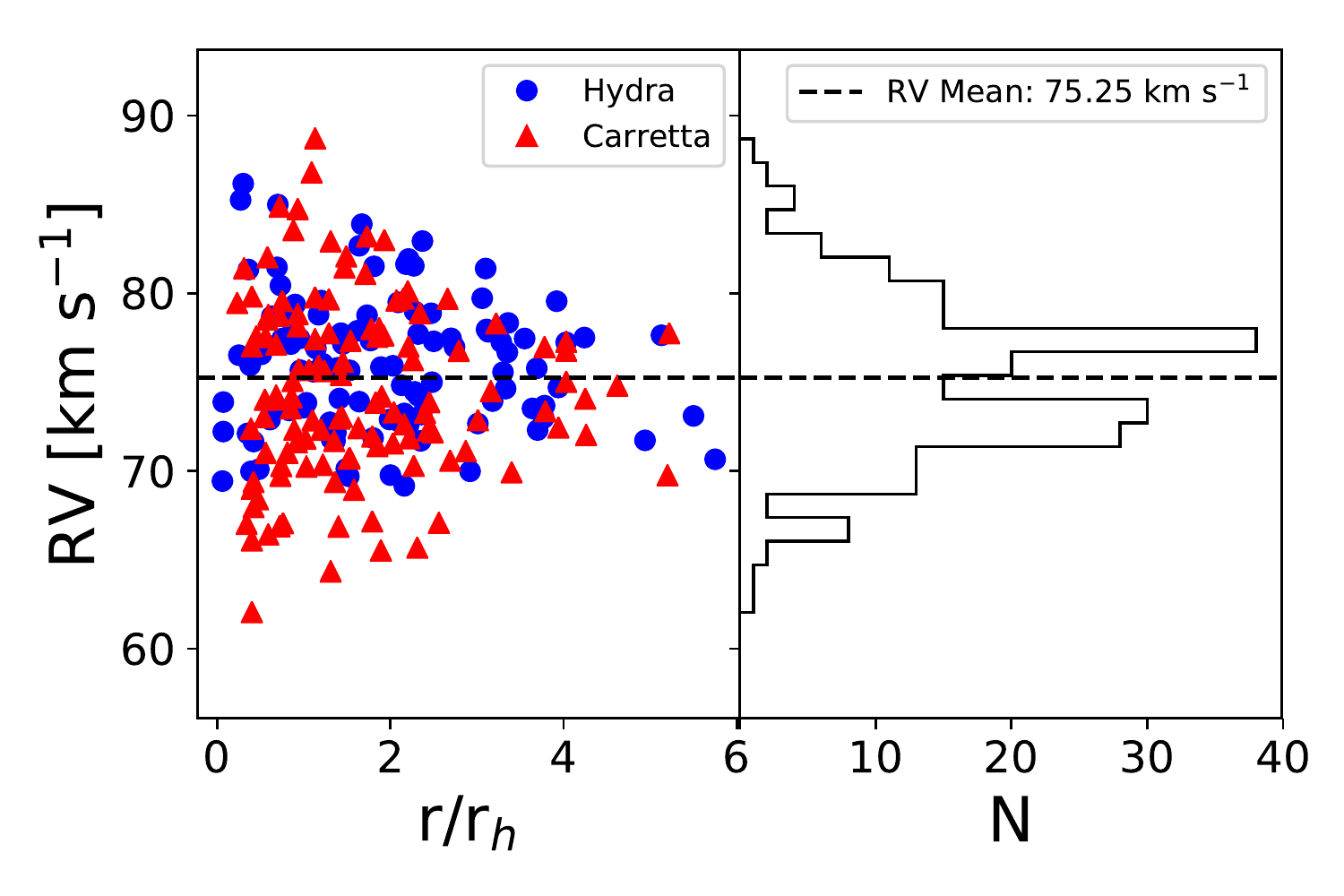}
\caption{RVs, including those from \citet{carrettaA}, of M10 as a function of radial distance from the cluster center in terms of half-mass radius (r$_{\rm h}$ = 117\arcsec \citep{harris}). The dashed line represents the mean of the entire sample. \citet{carrettaA} values have been adjusted for the offset between the studies as described in the text.}
\label{m10rRV}
\end{figure}

\begin{figure}
\centering
\includegraphics[trim = 0.4cm 0.4cm 0.4cm 0.4cm, scale=0.55, clip=False]{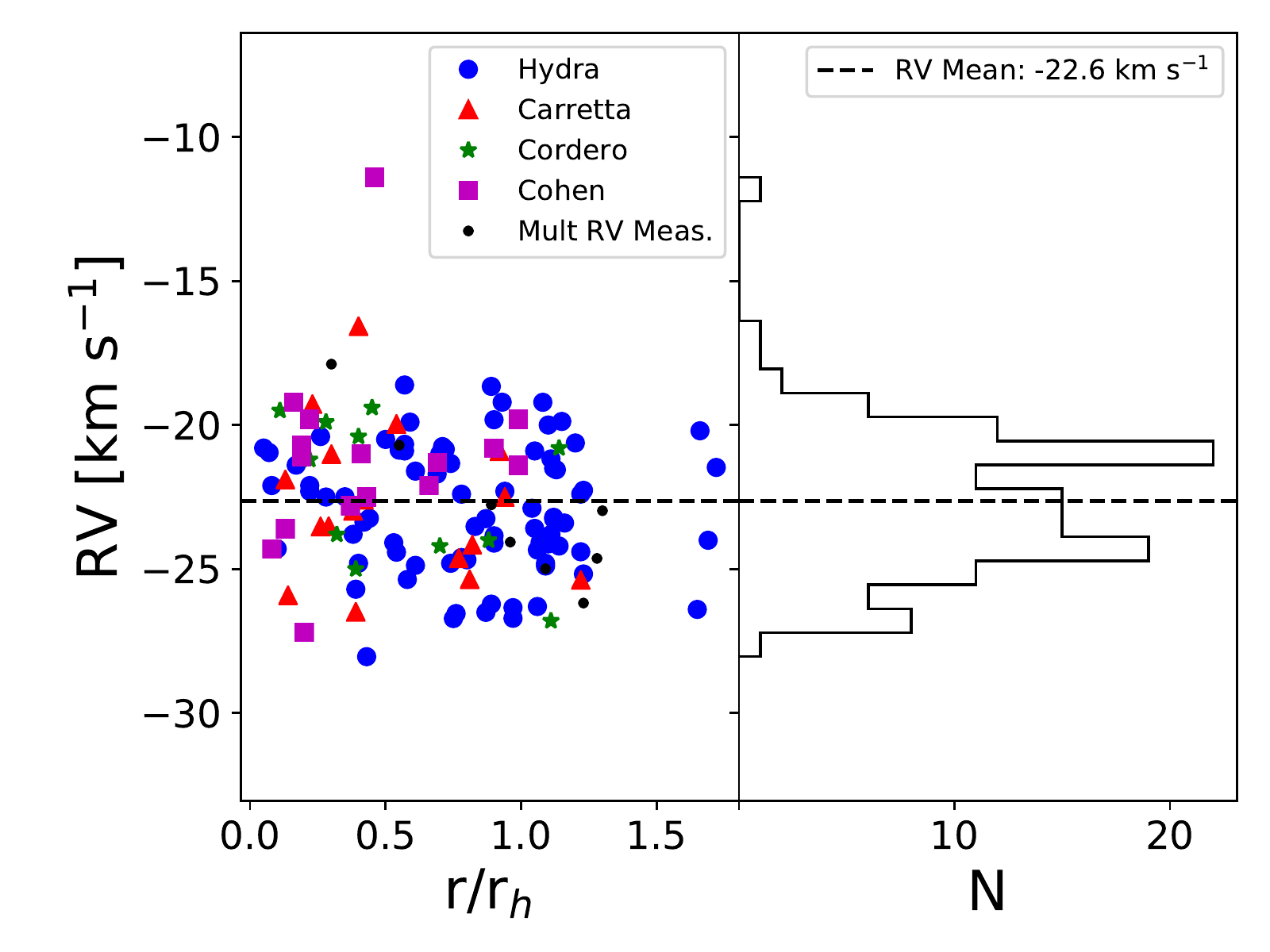}
\caption{As in Figure \ref{m10rRV} but for M71, the radial velocity measurements are plotted as a function of half mass radii (r$_{\rm h}$ = 146.2\arcsec \citep{cadelano}) from the cluster center. The dashed line represents the mean cluster velocity. Stars observed with Hydra are denoted as blue circles.  Stars with measurements from \citet{carrettaA}, \citet{cordero2015}, and \citet{cohen} are indicated as in the legend. If a star has multiple RV measurements in the literature, with no Hydra measurement, then it is plotted as a black dot and is the average of the multiple values.}
\label{m71rRV}
\end{figure}

By including the radial velocity measurements of 161 stars in M10 from \citet{carrettaA} we were able to double the sample of stars in our rotational analysis of M10. The two samples are of comparable size, but Figure \ref{m10rRV} shows that the Hydra sample extends and provides relatively more stars in the outer regions of M10 than the \citet{carrettaA} data. Combining the samples allows us to study possible radial changes in rotation signatures.

\citet{carrettaA} find a mean cluster velocity of $73.8 \pm 4.96$ km s$^{-1}$, suggesting a systematic offset from our velocities, which give a mean of $75.9 \pm 4.0$ km s$^{-1}$. Multiple measurements of stars in common allowed us to assess any systematic offsets between the data sets. Between the two, 21 stars were found in common yielding an average difference of RV (Hydra$-$Carretta) = $1.06 \pm 1.07$ (s.d.) km s$^{-1}$. The standard error of the mean is 0.23 km s$^{-1}$ indicating an offset in velocities significant enough to adjust for. Since the data obtained from the Hydra spectrograph were our primary source, we adjusted the data from \citet{carrettaA} by 1.06 km s$^{-1}$ to put the data sets on a common velocity scale. In cases where multiple measurements exist, we used the Hydra measurements. After correcting for the multiples between the two data samples, we had a total of 109 stars from our Hydra measures, and 118 stars from \citet{carrettaA}. We used a combined, and now unique, set of 225 stars to perform our rotation analysis of M10.

\subsubsection{M71 Literature Values}

For M71, we used three literature data sets in order to maximize the number of stars for our analysis (\citealt{carrettaA}, \citealt{cordero2015}, \citealt{cohen}). From \citealt{carrettaA}, 36 stars in M71 had RV measurements with a mean of $-23.2 \pm 2.66$ km s$^{-1}$, suggesting there is no offset when compared to the Hydra mean RV of $-22.9 \pm 2.2$ km s$^{-1}$. \citet{cordero2015} measured velocities for 35 stars with a mean RV of $-22.5 \pm 2.57$ km s$^{-1}$. \citet{cohen} measured 25 stars with a mean RV of $-21.1 \pm 3.40$ km s$^{-1}$.

\begin{table}
	\centering
	\caption{\textbf{RV Multiple Measurements}. 1. Offsets for our Hydra data come from multiple measurements between the 2016 and 2017 observing runs.}
	\label{M10mult}
	\begin{tabular}{ccccc} 
		\hline
		\hline
		{\textbf{GC}} & {\textbf{Data Set}} & {\textbf{\# Stars in}} & {\textbf{$\Delta$$\langle$RV$\rangle$}} & {\textbf{$\sigma$}}\\
        & & {\textbf{Common}} & {(km s$^{-1}$)} & {(km s$^{-1}$)}\\
		\hline
		M10 & Hydra$^{1}$ & 3 & 0.7 & 1.6\\
        \hline
        M10 & Carretta & 21 & 1.06 & 1.07\\
        \hline
        M71 & Hydra$^{1}$ & 3 & 0.4 & 0.9\\
        \hline
        M71 & Carretta & 11 & 0.0 & 1.5\\
        \hline
        M71 & Cordero & 12 & -0.6 & 0.4\\
        \hline
        M71 & Cohen & 5 & 0.0 & 0.9\\
		\hline
	\end{tabular}
\end{table}

Stars in common between each of the studies allowed us to access possible systematic differences in velocity scales. There exist 11, 12, and 5 multiple measurements for stars between our data and \citet{carrettaA}, \citet{cordero2015}, and \citet{cohen} respectively. As can be seen in Table \ref{M10mult}, the averages of the differences in velocities between the stars in common are close to 0. For this reason, we decided to use the data from each set unaltered from their literature values. As we did for M10, we used Hydra values when available and, in the case of multiple literature values, we took an average of available sources. This process resulted in a total of 129 stars in the M71 sample. The literature samples of M71 are comparable to the Hydra sample up to $\sim$ 1r$_{\rm h}$, while the Hydra sample contributes many more measurements in the outer region, out to 1.7r$_{\rm h}$.

\section{Analysis and Results} \label{Analysis}

\subsection{Methods} \label{methods}

We used equation 8 from \citet{walker} as a log-likelihood function
to estimate the mean velocity $\langle u \rangle$ and
velocity dispersion $\sigma_{p}$ of a sample of stars
(similar to the method used by \cite{kacharov}).
The log-likelihood function from \cite{walker} is shown in Equation \ref{one}.

\begin{equation}
\mathrm{ln}(p) = -\frac{1}{2}\sum_{i=1}^{N}\mathrm{ln}(\sigma_{i}^{2} + \sigma_{p}^{2}) -
\frac{1}{2}\sum_{i=1}^{N}\frac{(v_{i}-\langle u \rangle)^{2}}{(\sigma_{i}^{2}
+\sigma_{p}^{2})} - \frac{N}{2}\mathrm{ln}(2\pi)
\label{one}
\end{equation}

In Equation \ref{one}, $v_{i}$ and $\sigma_{i}$ are the measured quantities
representing the individual RVs and their errors, respectively. In
order to estimate $\langle u \rangle$ and $\sigma_{p}$, we used the
Metropolis-Hastings algorithm \citep{hastings} to sample the values that
maximize Equation \ref{one}. Our final estimates for $\langle u \rangle$ and $\sigma_{p}$
were taken as the mean of the individual marginalized distribution of each parameter
resulting from the Metropolis-Hastings algorithm. The error on each parameter was estimated
as the standard deviation of each of their respective distributions.
This method allows us to take the measurement error, and radial velocity dispersion,
into account when determining these characteristics of a given sample.

Using the method outlined above, we measured the mean velocity on
either side of a line bisecting the cluster. These mean velocities were then
subtracted from one another to form the value $\Delta\langle$RV$\rangle$. To estimate the error
on this quantity we added the errors on the mean of the velocities on each side of the line in
quadrature.

The line is then rotated counter clockwise in steps of 20$^\circ$, where the $\Delta\langle$RV$\rangle$ is determined at each PA. The $\Delta\langle$RV$\rangle$ as a function of PA is fit by the following expression,

\begin{equation}
\Delta \langle RV \rangle = \textrm{A}_{rot} \textrm{sin}(\textrm{PA} + \phi)
\label{two}
\end{equation}

$\textrm{A}_{rot}$ is defined as the amplitude, while $\phi$ is the phase of the sine curve fit to the data, where the position angle, PA$_{0}$ = 270 - $\phi$, defines the orientation of the cluster's axis of rotation (see e.g. \citealt{boberg}). In the sine curve fit, PA$_{0}$ corresponds to the first minimum and the actual amplitude of rotation measure is $\textrm{A}_{rot}$/2 (hereafter V$_{rot}$). In the interpretation of the rotation amplitude, it is important to consider that the rotational velocity measured is  V$_{rot}$sin(\textit{i}) and with no knowledge of the inclination, \textit{i}, V$_{rot}$ provides only a lower limit to the actual rotational velocity.

\subsection{Rotational Velocity Analysis} \label{rsa}

A few studies have explored the dependence of the internal rotational velocity of globular clusters on  the distance from the clusters' centers; the typical rotational curves found in these studies are characterized by an inner portion of the profile in which the rotational velocity increases with the distance from the center, a peak in the rotational velocity at a distance of about 1-2 half-mass radii and a declining rotation in the cluster's outer regions (see e.g. \citealt{boberg}, \citealt{lanzoni}; see also \citealt{lanzoni2018b} for an interesting cluster characterized by solid-body rotation over the entire radial range studied from its center to about 4 half-mass radii).

The number of stars available for each cluster in our study allows only a coarse characterization of the radial variation of the cluster's internal rotation; specifically we have used the half-mass radius as the point separating the inner portion of the cluster from the outer. We found rotation curves for the stars within $r_h$, outside of $r_h$, and the cluster as a whole. For M10 we adopt $r_h$ = 117\arcsec \citep{harris}, and for M71, $r_h$ = 146\arcsec \citep{cadelano}.

\begin{figure*}
\centering
\includegraphics[trim = 0.25cm 0.25cm 0.25cm 0.25cm, scale=0.45, clip=False]{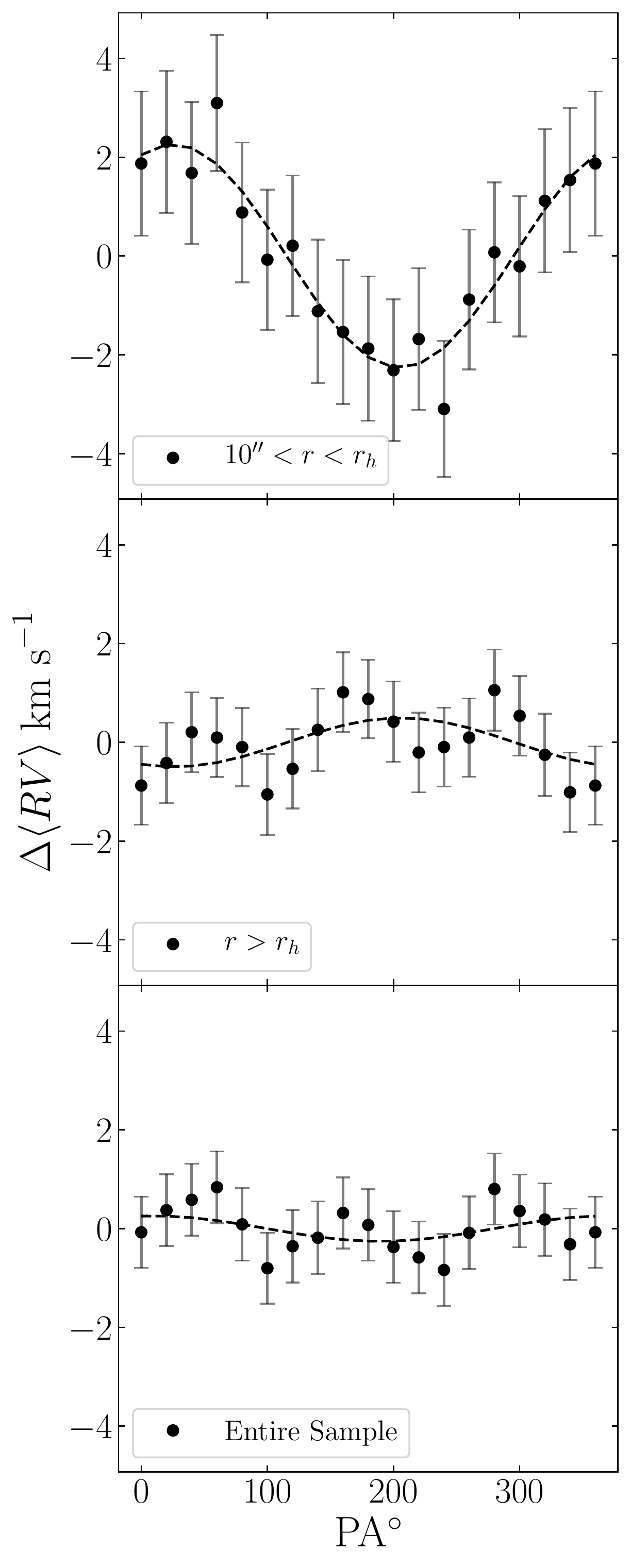}
\includegraphics[trim = 0.25cm 0.25cm 0.25cm 0.25cm, scale=0.45, clip=False]{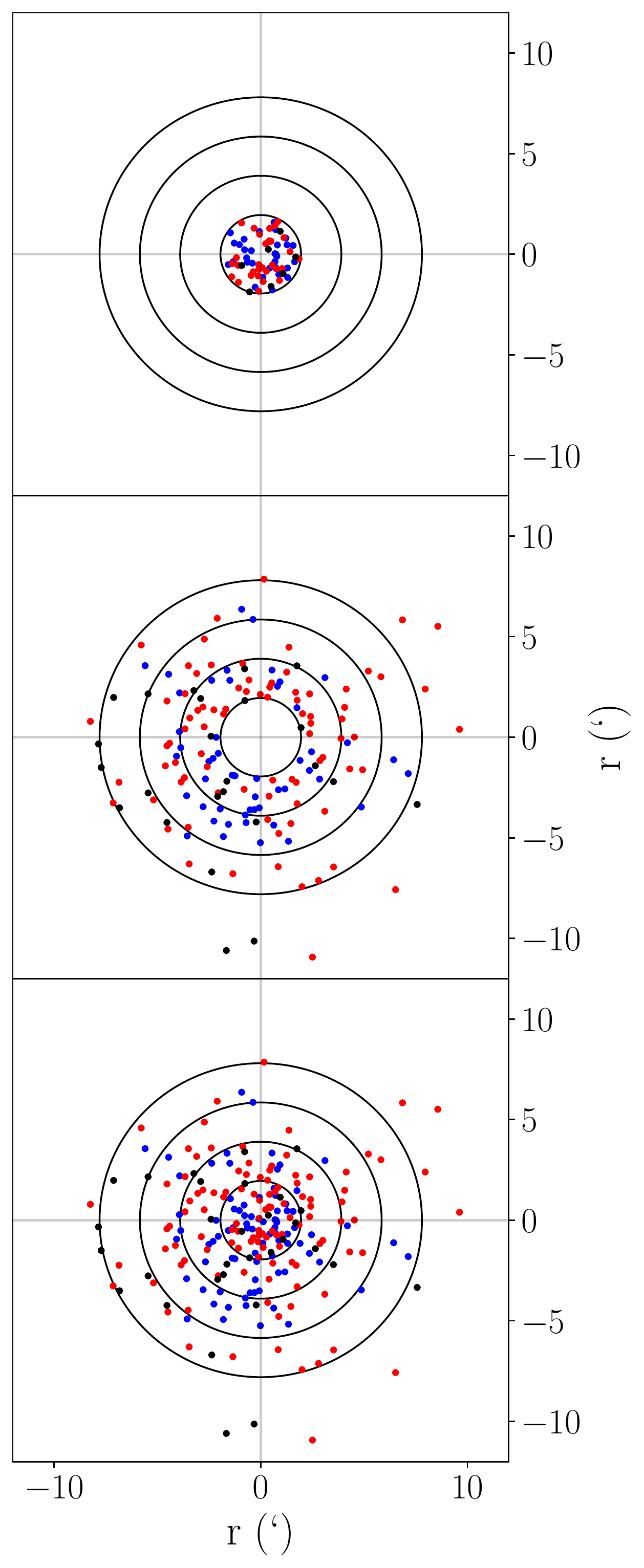}
\caption{Rotation curves and spatial distribution for the entire sample of M10 stars with RV measurements. The top panel refers to stars inside $r_h$ (117"), the middle panel refers to stars outside $r_h$, and the bottom panel refers to the entire sample. The black dots in the left panel show the $\Delta$$\langle$RV$\rangle$ as a function of PA in steps of 20$^\circ$. The dashed line is the sinusoidal fit to the data. The right panel shows the stellar distributions centered on the cluster center, with rings indicating increments of $r_h$. The blue dots on the distribution plots correspond to stars classified as first generation, the red dots to second generation stars (see section \ref{multPop}). The black dots correspond to measurements with no population data available.}
\label{m10rc}
\end{figure*}

In Figure \ref{m10rc} we show the rotation curves and spatial distribution of samples in the inner, outer, and entire cluster for M10. The rotation curves are fit to the sine curve of Equation \ref{two}, and the axis of rotation (PA$_{0}$) is taken as the first minimum of the curve. We find the maximum rotation signature in the inner region enclosed between 10\arcsec and the half-light radius.  In this region, the amplitude of the rotation curve is $\sim$2 km s$^{-1}$ at a PA$_{0}$ of 205$^\circ$.  No rotation is found in the innermost 10\arcsec (which include only three stars), in the outer region between r$_h$ and 5.8r$_h$, and for the entire sample at all radii. Table \ref{ressumM10} gives the results of the fits and analysis, with bin size used for each region in terms of half-mass radius and the number of stars used in each bin for M10. The errors listed for V$_{rot}$ in Table \ref{ressumM10} refer to the formal error in the fit that is calculated when fitting Equation \ref{two} to the data.

\begin{figure}
\centering
\includegraphics[scale=0.55, clip=False]{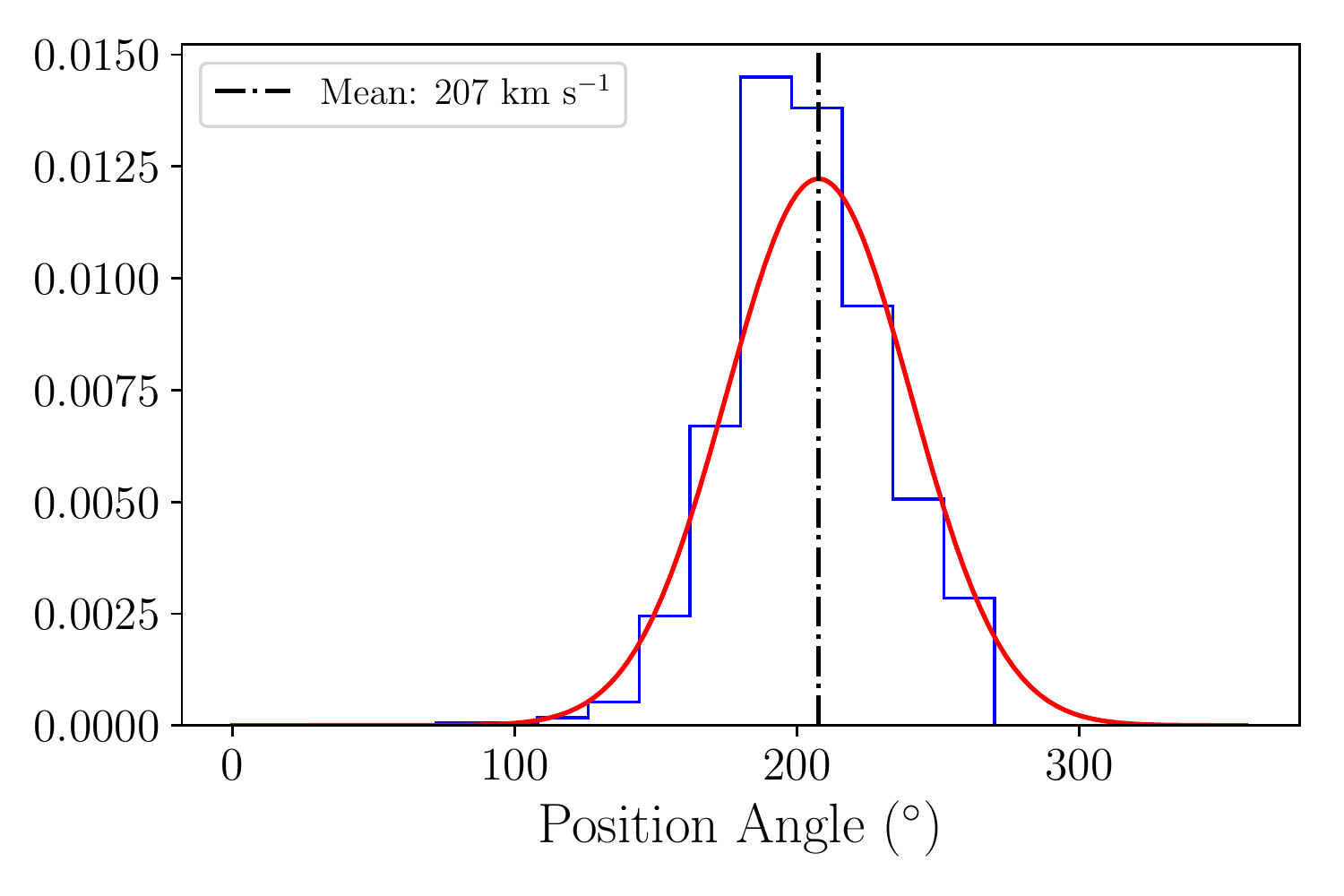}
\caption{Density histogram of 1000 PA$_{0}$ measurements from samples with replacement from the inner region of M10. The red line is a Gaussian with the mean (dot-dashed line) and dispersion from 1000 samples.}
\label{bootstrap}
\end{figure}

To assess the uncertainty in the orientation axis of rotation in the inner regions of M10, we randomly drew 1000 samples from our data with replacement to create subsamples with the same number of data points as our original set. We carried out the same rotational analysis on each of these subsamples to create a distribution of PA$_0$'s. Figure \ref{bootstrap} shows a density histogram of the sampling, from which a dispersion and error on the mean can be obtained for the average position angle found in the inner region of M10. Using this method, we found a mean PA$_{0}$ of 208$^{\circ}$, a dispersion of 34$^{\circ}$, and an error on the mean of 1.1$^{\circ}$. The mean PA$_{0}$ obtained using the bootstrap method agrees with the PA$_{0}$ of 205$^{\circ}$ found in Section \ref{rsa}. For the inner region of M10, we use this dispersion as our uncertainty in PA$_{0}$.

A  similar  rotation  amplitude  for  M10  was  found in the MIKiS study of globular cluster rotation \citep{ferraro}. Using 565 stars to perform a similar rotational analysis, they were able to determine an amplitude with confirming statistics as well. The PA$_0$ they measured is equal to 225$^{\circ}$ (if calculated with same choice of compass when assigning position  angles  as  in  this  paper).  While  close  to  our  measured value for the PA$_0$, they also found the highest V$_{\rm rot}$ of 1.4 km s$^{-1}$ in a region 190\arcsec$<$r$<$290\arcsec from the cluster center  as  opposed  to  our  highest V$_{\rm rot}$ in the  inner  region of 10\arcsec$<$r$<$117\arcsec. For the outer region of 190\arcsec$<$r$<$290\arcsec, \citet{ferraro} also listed poor t-student test statistics for M10, with $<90\%$ confidence interval of statistically different RVs on either side of their PA$_0$ consistent with our results.

\citet{kamann} found no obvious sign of rotation from a sample of over 9000 stars in the inner region of M10.  Their sample was limited to 0.8 r$_h$, and for the region of overlap with our sample inner region, showed V$_{\rm rot}$ values of $\sim$ 0.2 to 0.6 km s$^{-1}$. Their innermost two points, which are interior to the 10\arcsec\; limit of the Hydra data, showed some evidence of increasing rotation, with values of 1.5 to 2.5 km s$^{-1}$ but with larger uncertainties (see their figure A1). Their derived mean PA$_0$ of 143$^{\circ}$ (corresponding to 127$^{\circ}$ in our reference frame) reflects results dominated by the inner regions of the cluster that are not sampled by our data.

\begin{figure*}
\centering
\includegraphics[trim = 0.25cm 0.25cm 0.25cm 0.25cm, scale=0.45, clip=False]{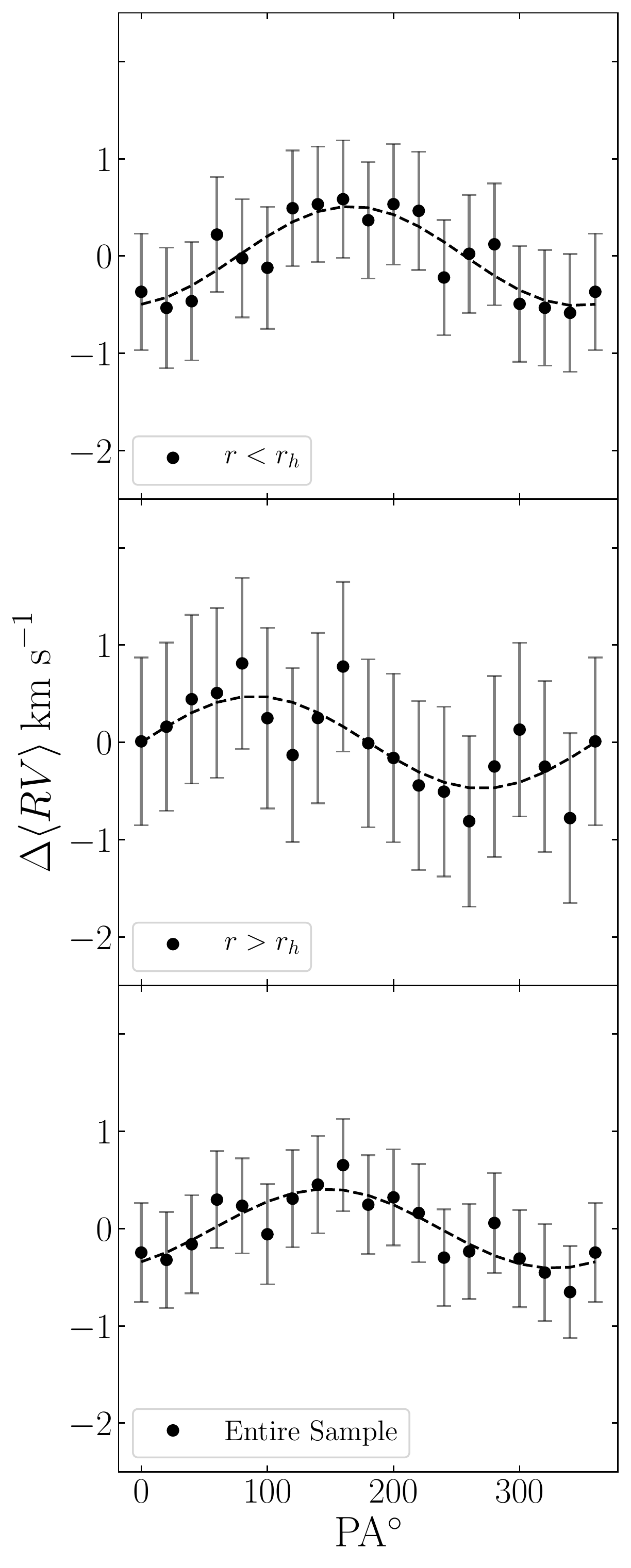}
\includegraphics[trim = 0.25cm 0.25cm 0.25cm 0.25cm, scale=0.45, clip=False]{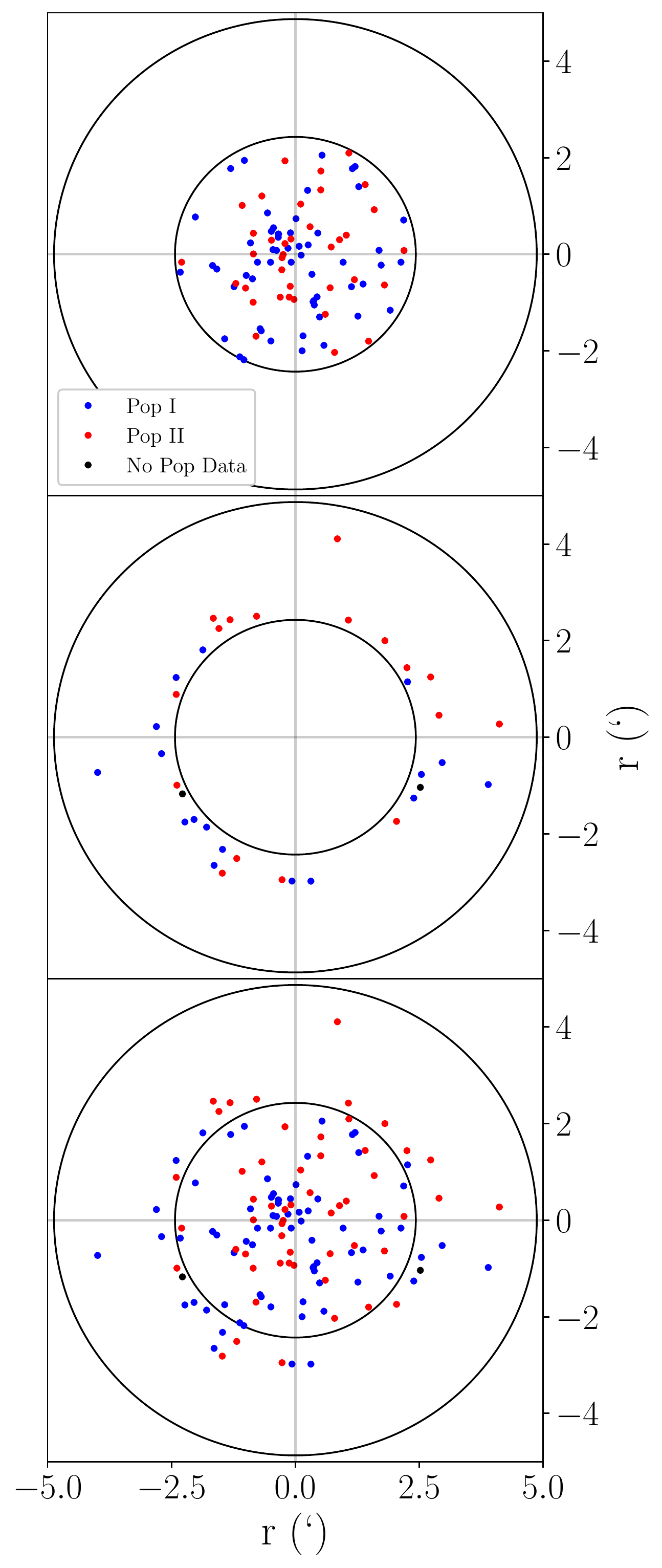}
\caption{Same as Figure \ref{m10rc}, but for M71. The half mass radius $r_h$ for M71 is 146\arcsec.}
\label{m71rc}
\end{figure*}

We show the rotation curves and spatial distribution for M71 in Figure \ref{m71rc}. Unlike M10, there are no regions within M71 that show any significant rotation. While the  data follow a sinusoidal variation, the low amplitudes and distribution of points do not reveal a significant rotation. Table \ref{ressumM71} outlines the results of the analysis for M71, where the region bin size, number of stars, position angle, amplitude of rotation, and the error to the fit are listed.

Possible  rotation in  M71 was  noted in \citet{cordero2015}.  \citet{kimmig} also completed a rotational analysis of M71 and reported a lack of rotation. The weak rotation signature of M71 could be a result of the effects of internal dynamical evolution and significant mass loss (see e.g. \citealt{cadelano} for possible evidence that M71 has lost a significant fraction of its initial mass).

\begin{table*}
	\centering
	\caption{\textbf{Rotational Analysis for M10}. \textbf{$r_i/r_h$} and \textbf{$r_f/r_h$} are the bin initial and final radii in terms of half-mass radius for the listed results. PA$_{0}$ is the position angle corresponding to the minimum of the rotation curve, PA$_{0} = 270 - \phi$. V$_{rot}$ is the actual maximum of the rotational velocity derived from fitting Equation \ref{two} to the rotation curves (see Section \ref{methods}).}
	\label{ressumM10}
	\begin{tabular}{cccccccc} 
		\hline
		{\textbf{r$_{i}$/r$_{\rm h}$}} & {\textbf{r$_{f}$/r$_{\rm h}$}} & {\textbf{N of Stars}} & {\textbf{PA$_{0}$}} & {\textbf{V$_{rot}$}} & {\textbf{V$_{rot}$ Error}} & {\textbf{KS}} & {\textbf{t-student}} \\
         & & & {($^{o}$)} & {(km s$^{-1}$)} & {(km s$^{-1}$)} & (\%) & (\%) \\
		\hline
		0.09 & 1 & 69 & 205 & 1.14 & 0.18 & 2.7 & 5.5\\
        \hline
        1 & 5.8 & 156 & 26 & 0.24 & 0.18 & 82.2 & 95.5\\
        \hline
        0 & 5.8 & 228 & 190 & 0.13 & 0.14 & 62.9 & 75.1\\
		\hline
	\end{tabular}
\end{table*}

\begin{table*}
	\centering
	\caption{\textbf{Rotational Analysis for M71}. \textbf{$r_i/r_h$} and \textbf{$r_f/r_h$} are the bin initial and final radii in terms of half-mass radius for the listed results. PA$_{0}$ is the position angle corresponding to the minimum of the rotation curve, PA$_{0} = 270 - \phi$. V$_{rot}$ is the actual maximum of the rotational velocity derived from fitting Equation \ref{two} to the rotation curves (see Section \ref{methods}).}
	\label{ressumM71}
	\begin{tabular}{cccccccc} 
		\hline
		{\textbf{r$_{i}$/r$_{\rm h}$}} & {\textbf{r$_{f}$/r$_{\rm h}$}} & {\textbf{N of Stars}} & {\textbf{PA$_{0}$}} & {\textbf{V$_{rot}$}} & {\textbf{V$_{rot}$ Error}} & {\textbf{KS}} & {\textbf{t-student}} \\
         & & & {($^{o}$)} & {(km s$^{-1}$)} & {(km s$^{-1}$)} & (\%) & (\%) \\
		\hline
		0 & 1 & 93 & 347 & 0.25 & 0.06 & 53.0 & 20.9\\
        \hline
        1 & 1.7 & 36 & 270 & 0.24 & 0.11 & 52.5 & 51.0\\
        \hline
        0 & 1.7 & 129 & 327 & 0.21 & 0.06 & 80.6 & 22.0\\
		\hline
	\end{tabular}
\end{table*}

\subsection{Statistical Significance of GC Rotation}\label{stats}

In order to evaluate the statistical significance  of the signature of rotation found in the previous section  we have carried out a t-student test and a Kolmogorov$-$Smirnov (KS) test. The t-student test establishes the significance of the difference between mean velocities of the two sets of RV measurements on opposite sides of PA$_{0}$ while the KS test gives the probability that the two sets are drawn from the same velocity distribution.

Tables \ref{ressumM10} and \ref{ressumM71} show the results of these tests for M10 and M71, respectively. Our tests show that the only statistically significant signature of rotation is that found in the inner regions of M10 (although the statistical significance is not very strong: p-value of the KS-test is 2.7\% and 5.5\% for the t-test). For M71, no region shows significant rotation; a larger sample size may be needed to establish the statistical significance of a weak rotation, if present.

\subsection{Velocity Dispersion Profiles}
\label{vdp}

In addition to searching for signatures of rotation within M10 and M71, we also constructed a velocity dispersion profile using the collected RVs. We split the sample of RVs into radial bins which contained a similar number of stars and fit the data to a Plummer model (see e.g. \citealt{heggie}). Within each radial bin we used the method described in section 3.1 to get an estimate of the mean velocity and velocity dispersion ($\sigma_{RV}$). This produces a measurement of the velocity dispersion in each radial bin while taking into account the mean velocity of the stars in the bin, their individual RV measurements, and errors on the RVs. The following equation was used to model the velocity dispersion as a function of radius:

\begin{equation}
\sigma^{2} = \frac{\sigma_{0}^{2}}{\sqrt{1 + \frac{r^{2}}{r_{\rm s}^{2}}}}
\label{equ2}
\end{equation}

\begin{figure}
    \centering
    \includegraphics[scale=0.6]{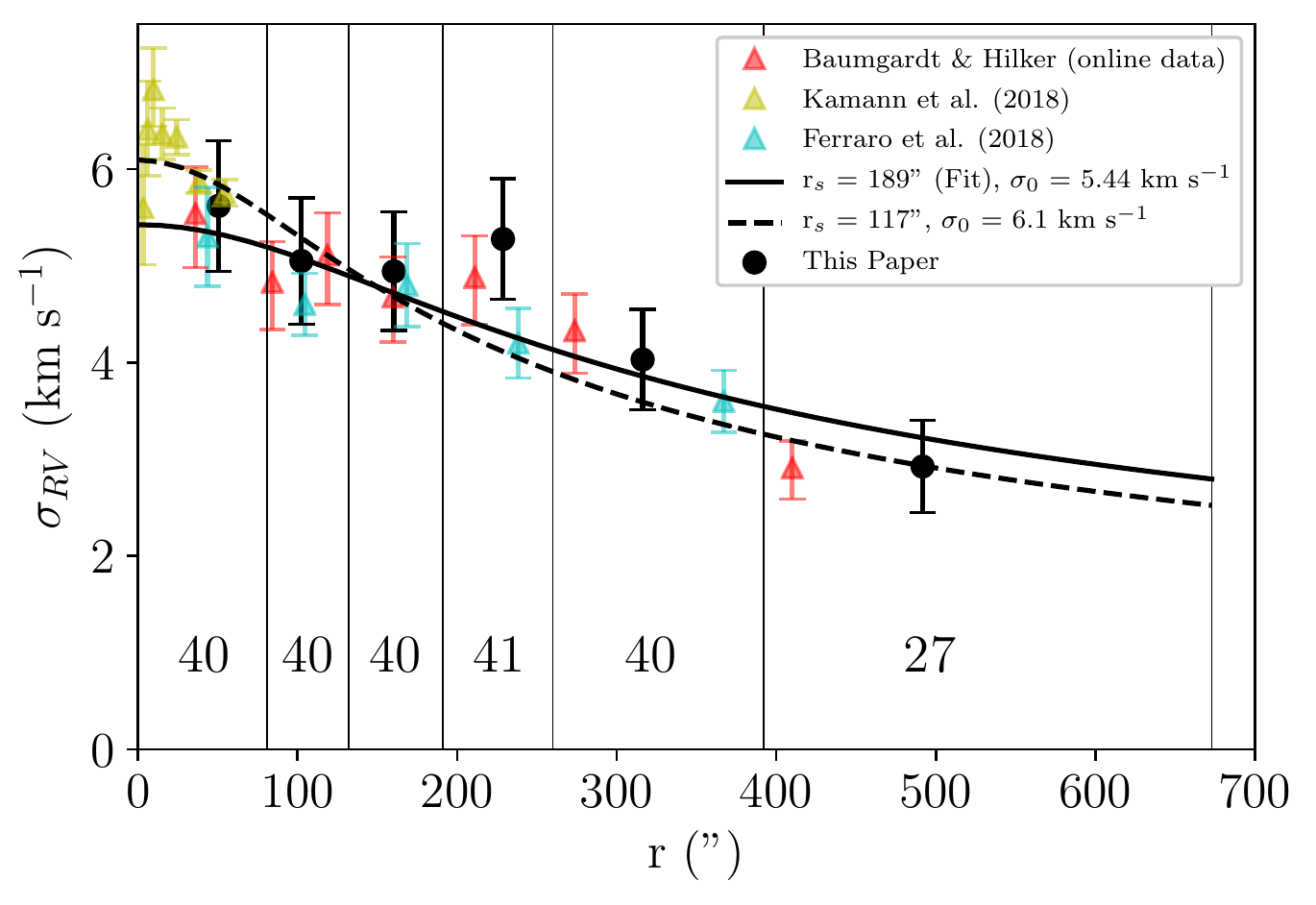}
    \caption{Radial profile of the velocity dispersion in M10. Black points indicate results from this paper; the solid black line represents a fit to these data performed with the scale radius (r$_{\rm s}$) and the central velocity dispersion ($\sigma_{0}$) as free parameters. The dashed line represents the fit with the scale radius fixed and equal to the observed projected half-mass radius (r$_{\rm h}$). The numbers along the bottom of the graph indicate how many RV measurements are in each bin. Triangles indicate data from the literature sources identified in the legend and discussed in the text.}
    \label{plummer}
\end{figure}

\begin{figure}
    \centering
    \includegraphics[scale=0.6, clip=False]{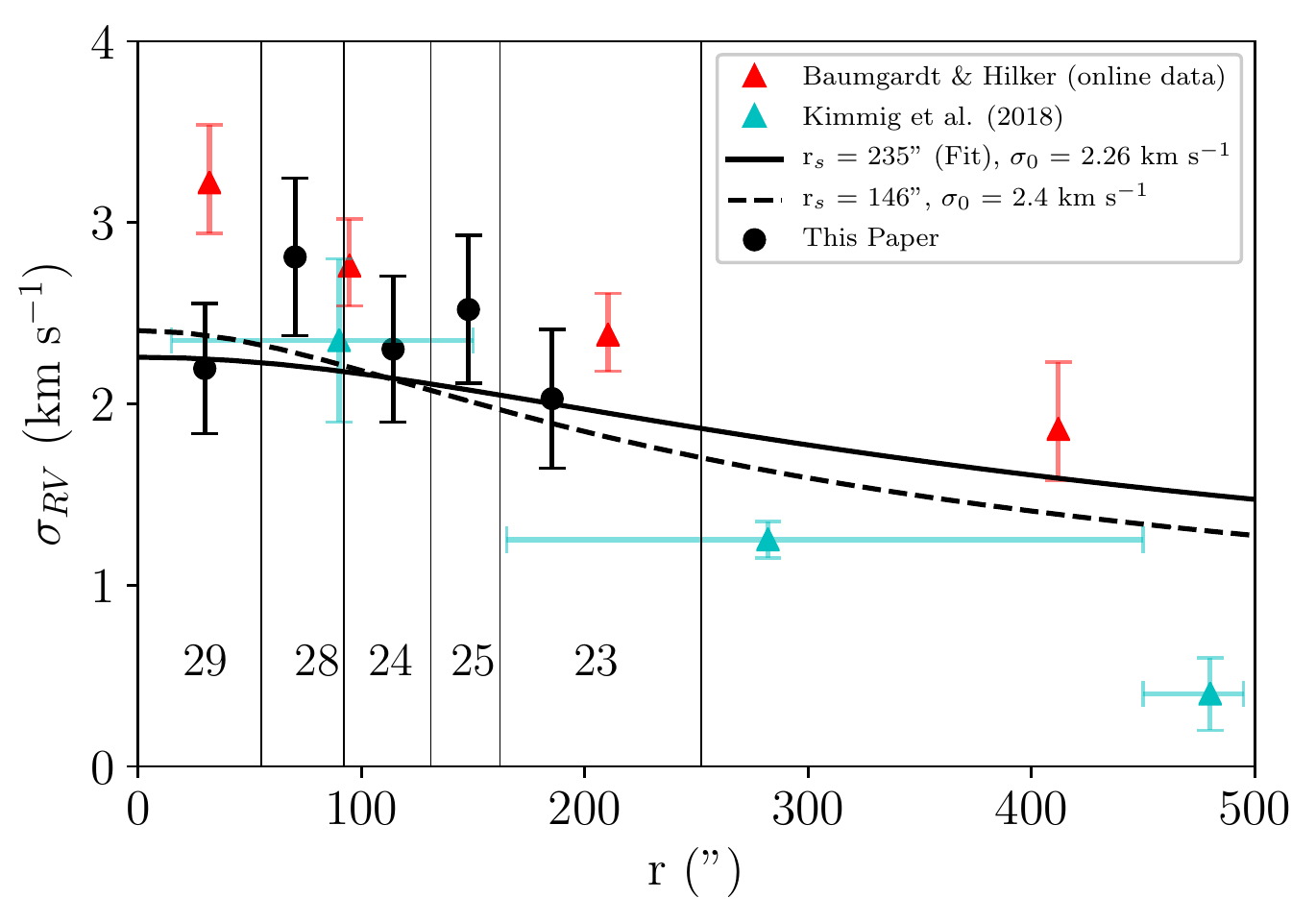}
    \caption{Same as Figure \ref{plummer}, but for M71.}
    \label{plummerm71}
\end{figure}

Where $\sigma_{0}$ is the central velocity dispersion and r$_{\rm s}$ is the scale radius. Figures \ref{plummer} and \ref{plummerm71} show the fit to our sample using Equation \ref{equ2} for M10 and M71, respectively. For each cluster, the fit was done twice, the first using $\sigma_{0}$ and r$_{\rm s}$ as free parameters and the second with r$_{\rm s}$ set equal to the observed projected half-mass radius, r$_{\rm h}$ (an assumption based on the fact that for a Plummer model the projected half-mass radius is equal to r$_{\rm s}$; see e.g. \citealt{heggie}).

In M10, the first fit resulted in a central velocity dispersion of $\sigma_{0} = 5.44 \pm 0.61$ km s$^{-1}$ and a scale radius of r$_{\rm s} = 189 \pm 24$\arcsec. When r$_{\rm s} =$ r$_{\rm h}$, the central velocity dispersion increased to $\sigma_{0} = 6.1 \pm 0.57$ km s$^{-1}$. A similar change in $\sigma_{0}$ and discrepancy among the scale radius found by the Plummer model fit and the recorded half-mass radius is noted in \citet{boberg}. These differences suggest more complex numerical or analytical models are needed to accurately describe these clusters.

Figure \ref{plummer} also shows data for the velocity dispersion for M10 from literature sources.  Our results agree very well in all cases, over the full range of radii sampled.  The Hydra, \citet{carrettaA}, and \citet{carrettaB} data provide an additional point at large distances that continues the decline in velocity dispersion seen in all studies.  In terms of central velocity dispersion, the values found here are in good agreement with those reported in the literature:  \citet{kamann} cite a central velocity dispersion of 5.6 $\pm$ 0.6 km s$^{-1}$, \citet{ferraro} find 6.0 $\pm$ 0.5 km s$^{-1}$, and \citet{baumgardt}\footnote{In both Figures \ref{plummer} and Figure \ref{plummerm71}, we show the updated values reported on the website https://people.smp.uq.edu.au/HolgerBaumgardt/globular/veldis.html} find a value of 6.2 km s$^{-1}$ from direct radial velocities and N$-$body simulations.

For M71, as seen in Figure \ref{plummerm71}, the velocity dispersion was much lower on average than M10, and with little variation produced a flatter fit. With both $\sigma_{0}$ and r$_{\rm s}$ as free parameters, the fit produced a velocity dispersion of $\sigma_{0} = 2.26 \pm 0.37$ km s$^{-1}$ and a scale radius of r$_{\rm s} = 235 \pm 39$\arcsec. Figure \ref{plummerm71} also shows literature values from \citet{kimmig} and \citet{baumgardt}. We see that the \citet{kimmig} data extends to significantly larger distances. In spite of the generally good agreement over the region of overlap, because of the steeply declining profile in the outer region, the \citet{kimmig} estimate of $\sigma_{0}$ is somewhat higher than what we find, at 3.1 $\pm 0.5$ km s$^{-1}$.

Data from \citet{baumgardt} show a clear decline in velocity dispersion with radius, but at values that are slightly high relative to the other studies, particularly at the innermost and outermost bins. We note that a difference in membership selection criteria may explain some of these differences. We applied a fairly strict criterion in proper motion membership (probability $>70$\% based on \citealt{cudworth} and private communication); less strict criteria applied by \citet{baumgardt} might result in the inclusion of stars we would have eliminated, and elevate velocity dispersions as a result. Given these differences, it is not surprising that the central velocity dispersion of 3.0 km s$^{-1}$ (\citealt{baumgardt} and web updates), is somewhat higher than our determination of 2.3 km s$^{-1}$.

\subsection{Multiple Populations}
\label{multPop}

As discussed in the Introduction, the rotational properties of multiple populations can provide key information necessary to shed light on the formation and dynamical history of multiple-population clusters (see e.g. \citealt{bekki}, \citealt{henault}, \citealt{mastrobuono}, \citealt{mastrobuono2016}, \citealt{tiongco2019}). For this reason, in addition to searching for global rotation in M10 and M71, we also conducted a rotation analysis of the individual populations of each cluster. By comparing the rotation profiles of two separate stellar populations, we can test whether different rotation signatures for these populations could dilute the global rotation profile of the cluster. In the case of one population rotating with an observable strength, for example, the non-rotating population stars can dilute and obscure the signal of the rotating population.

Population membership for M10 was determined following the studies of \citet{gerber} and \citet{carrettaB}. For stars in our sample that were measured by \citet{gerber}, we followed their method of separating stars based on the strength of a CN feature present at $\sim$3880 $\angstrom$. Using the $\delta$CN index measurements from that work, stars with a $\delta$CN higher than 0 were identified as second generation stars, while others were identified as first generation. If a star was not in the \citet{gerber} sample, we used the [Na/Fe] abundance from \citet{carrettaA} and \citet{carrettaB} to classify it based on the dividing value of 0.0 dex. While using two different classification methods could result in possible systematic offsets, we feel confident in our method based on the results of \citet{gerber}, which showed that both classification methods agree well with one another.

Our analysis in M10 shows that the rotation signature for each population separately is consistent with the global rotation signature. However, for the inner region, the rotation detected is no longer statistically significant, as a result of the smaller samples of stars (only 31 in each population).

For M71, stars were classified as described above for M10, following the method from \citet{gerber2020}.  No significant rotation was detected in either population, consistent with the lack of rotation seen in the full sample. Our findings are summarized in Table \ref{ressumM10_pop} (for M10) and Table \ref{ressumM71_pop} (for M71).

\begin{table*}
	\centering
	\caption{\textbf{Rotational Analysis for the multiple populations of M10.} \textbf{$r_i/r_h$} and \textbf{$r_f/r_h$} are the bin initial and final radii in terms of half-mass radius for the listed results. PA$_{0}$ is the position angle corresponding to the minimum of the rotation curve, PA$_{0} = 270 - \phi$. V$_{rot}$ is the actual maximum of the rotational velocity derived from fitting Equation \ref{two} to the rotation curves (see Section \ref{methods}).}
	\label{ressumM10_pop}
	\begin{tabular}{ccccccccc} 
		\hline
		{\textbf{Population}} & {\textbf{r$_{i}$/r$_{\rm h}$}} & {\textbf{r$_{f}$/r$_{\rm h}$}} & {\textbf{N of Stars}} & {\textbf{PA$_{0}$}} & {\textbf{V$_{rot}$}} & {\textbf{V$_{rot}$ Error}} & {\textbf{KS}} & {\textbf{t-student}} \\
        & & & & {($^{o}$)} & {(km s$^{-1}$)} & {(km s$^{-1}$)} & (\%) & (\%) \\
		\hline
		1 & 0.09 & 1 & 31 & 225 & 1.11 & 0.10 & 14.4 & 16.6\\
        \hline
        1 & 1 & 5.8 & 50 & 69 & 0.27 & 0.36 & 33.8 & 52.8\\
        \hline
        1 & 0 & 5.8 & 83 & 205 & 0.21 & 0.22 & 38.5 & 73.8\\
		\hline
		2 & 0.09 & 1 & 31 & 184 & 1.14 & 0.18 & 38.5 & 27.4\\
        \hline
        2 & 1 & 5.8 & 82 & 326 & 0.27 & 0.29 & 23.5 & 24.7\\
        \hline
        2 & 0 & 5.8 & 114 & 209 & 0.20 & 0.15 & 28.7 & 37.5\\
		\hline
	\end{tabular}
\end{table*}

\begin{table*}
	\centering
	\caption{\textbf{Rotational Analysis for the multiple populations of M71.} \textbf{$r_i/r_h$} and \textbf{$r_f/r_h$} are the bin initial and final radii in terms of half-mass radius for the listed results. PA$_{0}$ is the position angle corresponding to the minimum of the rotation curve, PA$_{0} = 270 - \phi$. V$_{rot}$ is the actual maximum of the rotational velocity derived from fitting Equation \ref{two} to the rotation curves (see Section \ref{methods}).}
	\label{ressumM71_pop}
	\begin{tabular}{ccccccccc} 
		\hline
		{\textbf{Population}} & {\textbf{r$_{i}$/r$_{\rm h}$}} & {\textbf{r$_{f}$/r$_{\rm h}$}} & {\textbf{N of Stars}} & {\textbf{PA$_{0}$}} & {\textbf{V$_{rot}$}} & {\textbf{V$_{rot}$ Error}} & {\textbf{KS}} & {\textbf{t-student}} \\
        & & & & {($^{o}$)} & {(km s$^{-1}$)} & {(km s$^{-1}$)} & (\%) & (\%) \\
		\hline
		1 & 0 & 1 & 26 & 68 & 0.18 & 0.13 & 96.4 & 79.3\\
        \hline
        1 & 1 & 1.7 & 47 & 340 & 0.13 & 0.08 & 20.5 & 96.0\\
        \hline
        1 & 0 & 1.7 & 73 & 228 & 0.08 & 0.08 & 67.6 & 86.2\\
		\hline
		2 & 0 & 1 & 19 & 14 & 0.48 & 0.15 & 29.4 & 17.1\\
        \hline
        2 & 1 & 1.7 & 35 & 323 & 0.58 & 0.10 & 47.3 & 20.9\\
        \hline
        2 & 0 & 1.7 & 54 & 339 & 0.48 & 0.07 & 17.2 & 17.8\\
		\hline
	\end{tabular}
\end{table*}

\section{Conclusions and Discussion}
\label{Conclusions}

In this paper we have carried out a study of the internal kinematics of two Galactic globular clusters, M10 and M71. The sample of radial velocities we have used include 107 new velocity measurements for M10 and 78 new velocity measurements for M71  obtained as part of this study.  We combined our RV measurements with literature data from \citet{carrettaA} for M10, and data from \citet{carrettaA}, \citet{cordero2015}, and \citet{cohen} for M71.

In order to characterize the kinematics of the multiple stellar populations found in these clusters, we have focused our attention only on velocity measurements of stars for which the additional spectroscopic data needed to identify the population membership were available.

We find a non-zero rotation amplitude both in the inner and the outer regions of M10. However, the statistical tests performed indicate that the rotation found in the outer regions of this cluster is not statistically significant. For the inner region  (10\arcsec $<$ r $<$ 117\arcsec), we find a marginally significant rotation amplitude of V$_{\rm rot}$ = 1.14 $\pm$ 0.18 km s$^{-1}$ with a PA$_{0}$ of 205 $\pm$ 34$^{\circ}$. The rotation amplitude we find for M10 is a lower limit because of the unknown inclination of the cluster, where V$_{\rm rot}$ = $\tilde{\rm V}_{\rm rot}$ $\sin i$ and $\tilde{\rm V}_{\rm rot}$ is the actual cluster's inner rotation velocity.

Using a Plummer model fit, we determine a central velocity dispersion of $\sigma_0 = 5.44 \pm 0.5$ km s$^{-1}$; the ratio of the rotation detected in the inner regions of M10 to the central velocity dispersion is equal to V$_{\rm rot}$/$\sigma_0$ $\sim$ 0.21.
Our analysis reveals no significant differences between the rotation properties of the stellar populations hosted by M10.

Our analysis of RV measurements in M71 shows that the rotation signal found in the inner, outer, and entire sample is not statistically significant and also for this cluster we find no significant differences between the rotation properties of its stellar populations.

Studies of globular cluster evolution and the evolution of their rotation over time show that as a globular cluster evolves and loses mass, its initial rotation gradually decreases as a result of angular momentum redistribution and loss; initial rotations must have been larger than what is seen today. This suggests that both M10 and M71 have undergone significant dynamical evolution and lost most of their initial rotation. The lack of any statistically significant difference between the rotation properties of the multiple populations of M10 and M71 is also likely to be due to the effects of dynamical evolution erasing any primordial difference in the kinematic properties of multiple populations. While statistical significance declines because of the reduced sample size, the amplitude and axis of rotation from the fit for each population is consistent with the global rotation properties. Should any memory of initial primordial differences in the kinematic properties of multiple populations be still present in these clusters, the identification of these differences will require larger sample sizes.

\section{Acknowledgements}
\label{acknowledgements}

We thank the anonymous referee for suggestions that significantly improved the paper.  We also thank Holger Baumgardt for information he provided on the M71 analysis.





\begin{thebibliography}{}

\bibitem[Baumgardt \& Hilker (2018)]{baumgardt} Baumgardt, H. \& Hilker, M.\ 2018, \mnras, 478, 1520

\bibitem[Baumgardt et al.(2019)]{baumgardt2019} Baumgardt, H.,  Hilker, M., Sollima, A., et al.\ 2019, \mnras, 482, 5138

\bibitem[Bellazzini et al.(2012)]{bellazzini} Bellazzini, M., Bragaglia, A., Carretta, E., et al.\ 2012, \aap, 538, A18

\bibitem[Bellini et al.(2015)]{bellini} Bellini, A., Vesperini, E., Piotto, G., et al.\ 2015, \apjl, 810, L13

\bibitem[Bellini et al.(2018)]{bellini2018} Bellini, A., Libralato, M., Bedin, L., et al.\ 2018, \apj, 853, 86

\bibitem[Bianchini et al.(2013)]{bianchini} Bianchini, P., Varri, A., Bertin, G., \& Zocchi, A.\ 2013, \apj, 772, 67

\bibitem[Bianchini et al.(2018)]{bianchini2018} Bianchini, P., van der Marel, R., del Pino, A., et al. \ 2018, \mnras, 481, 2125

\bibitem[Bekki (2010)]{bekki} Bekki, K. \ 2010, \apjl, 724, L99

\bibitem[Bekki (2011)]{bekki2011} Bekki, K. \ 2011, \mnras, 412, 2241

\bibitem[Boberg et al.(2017)]{boberg} Boberg, O., Vesperini, E., Friel, E., Tiongco, A., \& Varri, A. \ 2017, \apj, 841, 114

\bibitem[Bowman et al.(2017)]{bowman} Bowman, W., Pilachowski, C., van Zee, L., et al. \ 2017, \aj, 154, 4

\bibitem[Cadelano et al.(2017)]{cadelano} Cadelano, M., Dalessandro, E., Ferraro, F., et al.\ 2017, \apj, 836, 170

\bibitem[Carretta et al.(2009a)]{carrettaA} Carretta, E., Bragaglia, A., Gratton, R. G., et al.\ 2009a, \aap, 505, 117

\bibitem[Carretta et al.(2009b)]{carrettaB} Carretta, E., Bragaglia, A., Gratton, R. G., et al.\ 2009b, \aap, 505, 139

\bibitem[Chen et al.(2000)]{chen} Chen, L., Geffert, M., Wang, J., Reif, K. \& Braun, J. \ 2000, \aaps, 145, 223

\bibitem[Cohen et al.(2001)]{cohen} Cohen, J., Behr, B., \& Briley, M. \ 2001, \aj, 122, 1420

\bibitem[Cordero et al.(2015)]{cordero2015} Cordero, M., Pilachowski, C., Johnson, C., \& Vesperini, E.\ 2015, \apj, 800, 3

\bibitem[Cordero et al.(2017)]{cordero2017} Cordero, M., Henault-Brunet, V., Pilachowski, C., Balbinot, E., Johnson, C., Varri, A.\ 2017, \mnras, 465, 3515

\bibitem[Cordoni et al.(2020)]{cordoni} Cordoni, G., Milone, A., Mastrobuono-Battisti, A., et al.\ 2020, \apj, 889, 18

\bibitem[Cote et al.(1994)]{cote1994} Cote, P., Welch, D., Fischer P., \& Da Costa, G.\ 1994, \apjs, 90, 83

\bibitem[Cote et al.(1995)]{cote} Cote, P., Welch, D., Fischer P., \& Gebhardt, K.\ 1995, \apj, 454, 788

\bibitem[Cudworth (1985)]{cudworth} Cudworth, K. \ 1985, \aj, 90, 65

\bibitem[Dalessandro et al.(2018)]{dalessandro} Dalessandro, E., Mucciarelli, A., Bellazzini, M., et al.\ 2018, \apj, 864, 33

\bibitem[Einsel \& Spurzem, (1999)]{einsel} Einsel, C. \& Spurzem, R. \ 1999, \mnras, 302, 81

\bibitem[Ernst et al.(2007)]{ernst} Ernst, A., Glaschke, P., Fiestas, J., et al.\ 2007, \mnras, 377, 465

\bibitem[Fabricius et al. (2014)]{fabricius} Fabricius, M., Noyola, E., Rukdee, S., et al.\ 2014, \apjl, 787, L26

\bibitem[Ferraro et al.(2018)]{ferraro} Ferraro, F., Mucciarelli, A., Lanzoni, B., et al.\ 2018, \apj, 860, 50

\bibitem[Gaia Collaboration (2018)]{gaia} Gaia Collaboration, Helmi, A., van Leeuwen, F. et al.\ 2018, \aap, 616, A12

\bibitem[Gerber et al.(2018)]{gerber} Gerber, J., Friel, E., \& Vesperini, E. \ 2018, \aj, 156, 6

\bibitem[Gerber et al.(2020)]{gerber2020} Gerber, J., Friel, E., \& Vesperini, E. \ 2020, \aj, 159, 2

\bibitem[Gratton et al.(2012)]{gratton} Gratton, R., Carretta, E., \& Bragaglia, A. \ 2012, \aapr, 20, 50

\bibitem[Gratton et al.(2019)]{gratton2019} Gratton, R., Bragaglia, A., Carretta, E., et al. \ 2019, \aapr, 27, 1

\bibitem[Harris, (1996, 2010 Edition)]{harris} Harris, W.\ 1996, \aj, 112, 1487

\bibitem[Hastings, (1970)]{hastings} Hastings, W., \ 1970, Biometrika, 57, 97

\bibitem[Heggie \& Hut, (2003)]{heggie} Heggie, D. \& Hut, P. \ 2003, The Gravitational Million-Body Problem: A Multidisciplinary Approach to Star Cluster Dynamics

\bibitem[Henault-Brunet et al.(2015)]{henault} Henault-Brunet, V., Gieles, M., Agertz, O., et al. \ 2015, \mnras, 450, 2

\bibitem[Hong et al.(2013)]{hong} Hong, J., Kim, E., Lee, H., et al.\ 2013, \mnras, 430, 2960

\bibitem[Jindal et al.(2019)]{jindal} Jindal, A., Webb, J., \& Bovy, J. \ 2019, \mnras, 487, 3

\bibitem[Kacharov et al.(2014)]{kacharov} Kacharov, N., Bianchini, P., Koch, A., et al.\ 2014, \aap, 567, A69

\bibitem[Kamann et al.(2018)]{kamann} Kamann, S., Husser, T., Dreizler, S., et al.\ 2018, \mnras, 473, 5591

\bibitem[Kimmig et al.(2015)]{kimmig} Kimmig, B., Seth, A., Ivans, I., et al.\ 2015, \aj, 149, 53

\bibitem[Lane et al.(2009)]{lane2009} Lane, R., Kiss, L., Ibata, R., et al.\ 2009, \mnras, 400, 917

\bibitem[Lane et al.(2010b)]{lane2010b} Lane, R., Brewer, B., Kiss, L., et al.\ 2010, \apjl, 711, L122

\bibitem[Lane et al.(2011)]{lane2011} Lane, R., Kiss, L., Lewis, G., et al.\ 2011, \aapr, 530, A31

\bibitem[Lanzoni et al.(2018a)]{lanzoni} Lanzoni, B., Ferraro, F., Mucciarelli, A., et al.\ 2018a, \apj, 861, 16

\bibitem[Lanzoni et al.(2018b)]{lanzoni2018b} Lanzoni, B., Ferraro, F., Mucciarelli, A., et al.\ 2018b, \apj, 865, 11

\bibitem[Libralato et al.(2018)]{libralato} Libralato, M., Bellini, A., van der Marel, R., et al.\ 2018, \apj, 861, 99

\bibitem[Mastrobuono-Battisti \& Perets (2013)]{mastrobuono} Mastrobuono-Battisti, A, \& Perets, H. \ 2013, \apj, 779, 85

\bibitem[Mastrobuono-Battisti \& Perets (2016)]{mastrobuono2016} Mastrobuono-Battisti, A, \& Perets, H. \ 2016, \apj, 823, 61

\bibitem[Milone et al.(2018)]{milone} Milone, A., Marino, A., Renzini, A., et al.\ 2018, \mnras, 481, 5098

\bibitem[Pollard et al.(2005)]{photometry} Pollard, D.L., Sandquist, E.L., Hargis, J.R., \& Bolte, M.\ 2005, \apj, 628, 729

\bibitem[Reijns et al.(2006)]{reijns} Reigns, R., Seitzer, P., Freeman, K., et al. \ 2006, \aap, 445, 503

\bibitem[Richer et al.(2013)]{richer} Richer, H.B., Heyl, J., Anderson, J., et al.\ 2013, \apj, 771, L15

\bibitem[Sollima et al.(2019)]{sollima} Sollima, A, Baumgardt, H., \& Hilker, M.\ 2019, \mnras, 485, 1460

\bibitem[Tiongco et al.(2016)]{tiongco2016} Tiongco, M., Vesperini, E., \& Varri, A.\ 2016, \mnras, 455, 3693

\bibitem[Tiongco et al.(2017)]{tiongco} Tiongco, M., Vesperini, E., \& Varri, A.\ 2017, \mnras, 469, 683

\bibitem[Tiongco et al.(2019)]{tiongco2019} Tiongco, M., Vesperini, E., \& Varri, A.\ 2019, \mnras, 487, 4

\bibitem[Walker et al.(2006)]{walker} Walker, M.,  Mateo, M., Olszewski, E., et al.\ 2006, \aj, 131, 2114

\bibitem[Watkins et al.(2015)]{watkins} Watkins, L., van der Marel, R., \& Bellini, A., et al.\ 2015, \apj, 803, 29

\end{thebibliography}





\bsp	
\label{lastpage}
\end{document}